\documentclass[aps,
prx,
superscriptaddress,
twocolumn,
citeautoscript]{revtex4-2}

\newcommand{\AMP}{ICTPM}

\usepackage{float}
\usepackage{amsmath}
\usepackage{amssymb}
\usepackage{amsfonts}
\usepackage{dsfont}
\usepackage{graphicx,graphics}
\usepackage{color}
\usepackage[autostyle]{csquotes}
\usepackage{svg}
\usepackage{changes}

\newcommand{\ket}[1]{\lvert #1\rangle}
\newcommand{\bra}[1]{\langle#1 \rvert}

\newcommand{\ex}[1]{\langle #1\rangle}

\newcommand{\ann}{\hat{a}}
\newcommand{\adag}[1][]{\hat{a}^{\dagger #1}}

\newcommand{\bnn}{\hat{b}}
\newcommand{\bdag}[1][]{\hat{b}^{\dagger #1}}

\newcommand{\cnn}{\hat{c}}
\newcommand{\cdag}[1][]{\hat{c}^{\dagger #1}}

\newcommand{\affDLR}[0]{German Aerospace Center (DLR), Institute of Quantum Technologies, 89081 Ulm, Germany}
\newcommand{\affUniUlm}[0]{Institute for Complex Quantum Systems and IQST, University of Ulm, 89069 Ulm, Germany}
\newcommand{\affSherbrooke}[0]{Institut Quantique, Universit\'e de Sherbrooke, Sherbrooke, Qu\'ebec J1K 2R1, Canada}

\begin{document}

\title{Amplification and Detection of Single Itinerant Microwave Photons}

\author{Lukas Danner}
\email{lukas.danner@dlr.de}
\affiliation{\affDLR}
\affiliation{\affUniUlm}

\author{Max Hofheinz}
\affiliation{\affSherbrooke}

\author{Nicolas Bourlet}
\affiliation{\affSherbrooke}

\author{Ciprian Padurariu}
\affiliation{\affUniUlm}

\author{Joachim Ankerhold}
\affiliation{\affUniUlm}

\author{Bj\"orn Kubala}
\affiliation{\affDLR}
\affiliation{\affUniUlm}

\date{\today}
\begin{abstract}
Single-photon detectors are an essential part of the toolbox of modern quantum optics for implementing quantum technologies and enabling tests of fundamental physics. The low energy of microwave photons, the natural signal path for superconducting quantum devices, makes their detection much harder than for visible light. Despite impressive progress in recent years and the proposal and realization of a number of different detector architectures, the reliable detection of a single itinerant microwave photon remains an open topic.

Here, we investigate and simulate a detailed protocol for single-photon multiplication and subsequent amplification and detection. At its heart lies a Josephson-photonics device
which uses inelastic Cooper-pair tunneling driven by a dc bias in combination with the energy of an incoming photon to create multiple photons, thus compensating for the low-energy problem. 
Our analysis provides clear design guidelines for utilizing such devices, which have previously been operated in an amplifier mode with a continuous wave input, for counting photons. 
Combining a formalism recently developed by M{\o}lmer to describe the full quantum state of in- and outgoing photon pulses with stochastic Schr\"odinger equations, we can describe the full multiplication and detection protocol and calculate performance parameters, such as detection probabilities and dark count rates.
With optimized parameters, a high population of a single output mode can be achieved that can then be easily distinguished from vacuum noise in heterodyne measurements of quadratures with a conventional linear amplifier. 
Realistic devices with two multiplication stages with multiplication of $16$ reach for an impinging Gaussian pulse of length $T$ a detection probability of $84.5\%$ with a dark count rate of $10^{-3}/T$, and promise to outperform competing schemes.

\end{abstract}
\maketitle

\section{Introduction}
\label{sec::Introduction}
The ability to reliably detect radiation on the single-photon level is an indispensable ingredient for numerous protocols and technologies in quantum information processing, quantum sensing, and quantum communication. Devices such as avalanche photodiodes and superconducting nanowire single-photon detectors \cite{You2020, Zadeh2021} provide this capability at frequencies above infrared, thus enabling quantum key distribution, ranging, heralded preparation of non-Gaussian states, boson sampling, and many more \cite{chen2021integrated,guan2022lidar,neergaard2006generation,sonoyama2024generation,ercolano2025photon,zhong2020quantum}.
However, other quantum technological devices, most prominently those based on superconducting circuits, naturally operate on energy scales compatible with microwave photons so that their efficient amplification, detection, and characterization is highly requested.

To meet the challenge of detecting itinerant microwave photons on the single-photon level, a considerable variety of schemes has been proposed previously \cite{romero2009photodetection,peropadre2011approaching,helmer2009quantum,iakoupov2020sequential,navez2023quantum,Royer_PRL_2018,he2024nonreciprocity}.
After a number of breakthrough experimental realizations\cite{inomata2016single,besse2018single,kono2018quantum}, the last decade has seen impressive progress in the field, see e.g.~brief reviews in \cite{casariego2023propagating,gu2017microwave,sathyamoorthy2016detecting}.
Among the schemes are threshold detectors, where photon absorption switches a current biased Josephson junction to the resistive state \cite{chen2011microwave,ladeynov2025detection,pankratov2022towards,rettaroli2021josephson,Oelsner_PRApp_2017,he2025experimental} or induces a phase transition in a Josephson parametric amplifier \cite{wang2025observing}, and various ways to capture an incoming photon in a cavity \cite{inomata2016single,besse2018single,kono2018quantum,lescanne2020irreversible,balembois2024cyclically}, where it interacts with a local system whose modified state is subsequently detected. 
Demonstrated applications include the improved sensing of signals from electron spin resonance \cite{albertinale2021detecting,billaud2025electron}, the detection of Josephson radiation\cite{karimi2024bolometric}, the measurement-based remote entanglement of superconducting qubits \cite{narla2016robust} as well as their readout \cite{opremcak2018measurement,gunyho2024single}, and, perhaps most prominently in recent years, in the fundamental physics quest for detecting axions \cite{Dixit_PRL_2021,braggio2025quantum,pankratov2022towards}. 

However, these realizations also have their significant limitations. Those that operate in regimes of  maximal nonlinearity and non-unitarity, map all incoming non-zero
photon number states onto the same output state. Further, they are typically gated or have a blind time after each detection event where the detector is reset which implies that these detectors saturate quickly. On the other hand, more linear power sensors either lack single photon sensitivity, as in the case of bolometers \cite{karimi2020quantum,karimi2024bolometric,kokkoniemi2019nanobolometer,kokkoniemi2020bolometer,gunyho2024single}, or achieve true number resolving single photon detection only for precisely known incoming modes sufficiently far apart in time \cite{dassonneville2020number}. 

A first step towards counting photons arriving at unknown times has recently been achieved in the form of a photon number amplifier based on inelastic Cooper pair tunneling \cite{Juha_ICTA_2018, Albert_ICTA_2024}. This Inelastic-Cooper-pair-Tunneling Photon Multiplier (\AMP) multiplies the photon number by an integer factor without added photon noise, at the expense of deamplifying or losing phase information. 
However, this fundamentally novel mode of amplification has so far only been investigated for classical (coherent continuous-wave) input. Here we theoretically analyze true quantum input (single-photon wavepackets) and describe how such a photon-number amplifier is then read out using a quantum-limited amplifier to detect and count individual photons.
The \AMP\ uses the energy $2eV_\mathrm{dc}$ of a Cooper pair tunneling inelastically through a Josephson junction biased at a voltage $2eV_\mathrm{dc} \approx n \hbar\omega_b - \hbar\omega_a$ to power a photon conversion process between one excitation from a microwave cavity $\ann$ to $n$ excitations in a cavity $\bnn$ which are both connected in series to the junction, see Fig.~\ref{fig:fig_model}. We utilize M{\o}lmer's approach \cite{ Kiilerich2019, Kiilerich2020, Christiansen2023,Khanahmadi2023, Lund2023} to describe a resonant Gaussian single-photon pulse impinging on the \AMP. While near-perfect photon conversion with negligible input-photon reflection can be achieved by adjusting the Josephson energy, a key challenge is that, because of the inherent nonlinearity of the photon number amplification process, a single incoming temporal mode may be mapped to multiple outgoing modes, each containing less than $n$ photons. We analyze and characterize different regimes of the output, identifying parameters where a single temporal output mode is highly occupied. Numerically, we employ quantum trajectories based on the stochastic Schr{\"o}dinger equation to carefully model the experimental readout of heterodyne quadrature measurement using a quantum-limited amplifier with large gain $G$. Thereby, we provide a scheme to analyze the output signal in order to detect the single input photon. Crucially, the detection efficiency can be optimized for parameter regimes exhibiting few output modes, where one is highly populated and can be distinguished from vacuum. 

However, achieving a large multiplication factor $n$ in order to improve detection efficiency would, in experiments where $\omega_b$ and $\omega_b$ are fixed, lead to a large dc-voltage. In this situation, the \AMP\ can access higher-order processes, resulting in emission without an incoming signal \cite{Albert_ICTA_2024} which, in turn, would trigger false detection events, i.e. dark counts. To avoid this limitation, we also analyze photon multiplication in a setup with two stages. Physically, internal processes here strongly differ from a single-stage setup, which strongly influences the system dynamics and the structure of the output signal.  

\section{Model Description and Simulations}
\label{sec::Model}

\begin{figure}[t]
    \centering
    \includegraphics[width=\columnwidth]{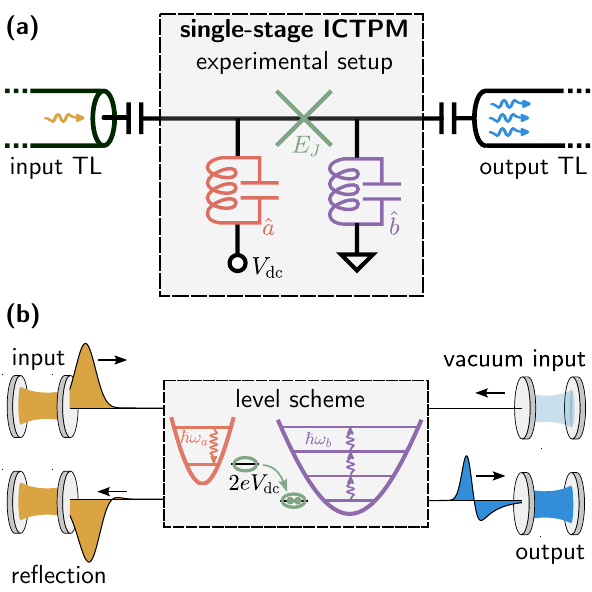}
    \caption{Inelastic-Cooper-pair-Tunneling Photon Multiplier. (a) The system consists of a dc-biased Josephson junction (with tunable Josephson energy $E_J$), connected in series with two microwave modes, that are realized as LC-resonators and capacitively coupled to transmission lines (TLs). A single-photon pulse impinging on cavity $\ann$ from the input transmission line will be absorbed by the \AMP, which then leaks $n$ photons into the output transmission line. (b) A Cooper pair can inelastically tunnel through the junction if the microwave modes absorb its energy of $2eV_\mathrm{dc}$. By setting the dc-voltage on the resonance $2eV_\mathrm{dc}/\hbar + \omega_a= n \omega _b $, the Cooper pairs drive the transition where one cavity-$\ann$ excitation is coherently transferred to $n \in \mathbb{N}$ cavity $\bnn$-excitations. Time-dependent input-, output- and reflected pulses moving along transmission lines which are absorbed or emitted from the \AMP\ can be numerically modeled by M{\o}lmer's approach \cite{Christiansen2023, Kiilerich2019, Kiilerich2020, Khanahmadi2023, Lund2023}. Therein, the transmission lines are mimicked by auxiliary cavities which emit (absorb) modes $u(t)$ ($v(t)$) for appropriately chosen time-dependent loss rates $g_u(t)$ ($g_v(t)$). 
    }
    \label{fig:fig_model}
\end{figure}

  We consider a single itinerant microwave photon in a superconducting input transmission line [cf. Fig.~\ref{fig:fig_model}(a)], which is the signal pulse to be detected. Before detection, the signal is amplified by an Inelastic-Cooper-Pair-Tunneling Photon Multiplier (\AMP). First, we introduce 
  the \AMP\ \cite{Juha_ICTA_2018,  Albert_ICTA_2024}, which consists of a dc-biased Josephson junction connected in series to two microwave cavities (realized as $LC$-oscillators). The \AMP\ is described by the Hamiltonian $\hat{H}_1 = \hbar \omega_a \adag \ann  + \hbar \omega_b \bdag \bnn + \hat{H}_J$, where two harmonic oscillators with eigenfrequencies $\omega_{\xi} = (L_{\xi} C_{\xi})^{-1/2}$ ($\xi=a,b$) are driven nonlinearly by the Josephson junction 
  \begin{equation}
        \label{eq::1stage_Hamiltonian}
      \hat{H}_J =  - E_J \cos\left[\frac{2e}{\hbar}V_\mathrm{dc} t + \alpha_0 (\adag + \ann) + \beta_0 (\bdag + \bnn) \right] 
 \end{equation}
  with driving energy $E_J$. Here, we have introduced the zero-point fluctuations $\alpha_0 = (\pi Z_a / R_Q)^{1/2}$ (and $\beta_0$ analogously) of the cavities determined by their characteristic impedance $Z_{\xi} = (L_{\xi} / C_{\xi})^{1/2}$ and the quantum of resistance, $R_Q = h/4e^2$. The phase argument of the Josephson junction then follows from a Kirchhoff's loop rule for the (time-integrated) voltages of the circuit shown in Fig.~\ref{fig:fig_model}(a).
 Both cavities are capacitively coupled to respective input or output transmission lines, leading to an effective loss rate $\gamma_{\xi}$, so that
the \AMP\ can be modeled by a Lindblad master equation with dissipation operators $\hat{L}_{\xi} = \sqrt{\gamma_\xi} \hat{\xi}$. 
The coherent coupling Hamiltonian, $\hat{H}_J$, describing Cooper-pair tunneling across the dc-biased Josephson junction, only becomes effective at certain resonances, where the energy provided by the voltage bias to a tunneling pair, $2eV_\mathrm{dc}$, can be exactly absorbed by changing the occupation of the electromagnetic modes of the circuit. Here, we tune the dc-voltage  to a resonance condition $2e V_\mathrm{dc}/\hbar +\omega_a\approx n \omega_b $ ($n \in \mathbb{N}$), such that each tunneling Cooper pair creates $n$ excitations in cavity $\bnn$ by annihilating one excitation from cavity $\ann$.

This photon process is indeed selected by a rotating wave approximation (RWA) of the coupling Hamiltonian~\eqref{eq::1stage_Hamiltonian} (see Appendix \ref{sec::app_1stage_RWA}), i.e.,  
  \begin{equation}
      \label{eq::1stage_Hamiltonian_RWA}
      \hat{H}_{J,\mathrm{RWA}} 
      \approx \frac{E_J^* \alpha_0 \beta_0^n}{2 n!} (\ann \bdag[n] + \adag \bnn^n)
  \end{equation}
  where $E_J^* = E_J e^{-(\alpha_0^2+\beta_0^2)/2}$ (see below for higher-order corrections to \eqref{eq::1stage_Hamiltonian_RWA}). 
  
  When an itinerant microwave photon in the input transmission line is resonantly absorbed by cavity $\ann$, the Josephson junction drives a multiplication 
  process, where one excitation from cavity $\ann$ is coherently transferred into $n$ excitations in cavity $\bnn$. These $n$ photons eventually leak into the output transmission line, where they can  be detected more easily. 
  
  Theoretically, the system's interaction with incoming and outgoing itinerant few-photon pulses can be conveniently modeled by a cascaded Master equation approach \cite{Christiansen2023, Kiilerich2019, Kiilerich2020, Khanahmadi2023, Lund2023}. Within input-output theory, one  defines an incoming temporal mode of the input field by the annihilation operator $\ann_{u} = \int_0^\infty u(t) \ann_{\mathrm{in}}(t) dt$ (where the mode envelope $u(t)$ is normalized, $\int_0^\infty |u(t)|^2 dt = 1$). This itinerant pulse is modeled by an auxiliary cavity coupled directly to the input cavity with time-dependent loss-rate $g_u(t) = u^*(t)/\sqrt{1 - \int_0^t |u(\tau)|^2 d\tau}$. 
  Such an auxiliary cavity will emit its initial quantum state $\hat{\rho}_u$ (where $\hat{\rho}_u$ should be chosen to be the quantum state of the mode $\ann_u$) into a temporal mode with envelope $u(t)$. Analogously, another auxiliary cavity with loss rate $g_v(t) = -v^*(t)/\sqrt{\int_0^t |v(\tau)|^2 d\tau}$ will perfectly absorb the reflected itinerant quantum pulse with mode envelope $v(t)$, where its final state $\hat{\rho}_v$ is the quantum state of the mode $\ann_v = \int_0^\infty v(t) \ann_{\mathrm{out}}(t) dt$. In the same manner, an auxiliary output cavity can be introduced for the right transmission line where we assume vacuum input. 
  The resulting cascaded Master equation (see Appendix \ref{sec::app_cascadedME}) then contains interaction terms between the auxiliary cavities and the system, as well as modified Lindblad operators $\hat{L}_a \rightarrow \sqrt{\gamma_a} \ann + g_u(t) \ann_u + g_v(t) \ann_v$ (and analogously for cavity $\bnn$). In that manner, the interaction of the \AMP\ with a small number of input and output pulses can be described. One big advantage of this method is that true quantum input, i.e. a single photon Fock state $\hat{\rho}_u = \ket{1_u}\bra{1_u}$ (which behaves inherently differently than simple coherent input), can be simulated. The second advantage is that this simulation is also more efficient: To describe a single-photon Fock state, we need $2$ states for the auxiliary input cavity while the amplifier dimension is then rigorously restricted to the dimension $2\times(n+1)$ within RWA, Eq.~(\ref{eq::1stage_Hamiltonian_RWA}). A coherent input with mean occupation of a $\langle \adag_u \ann_u =1 \rangle$ numerically requires a considerably larger Hilbert space dimension.

\begin{figure}[t]
    \centering
    \includegraphics[width=\columnwidth]{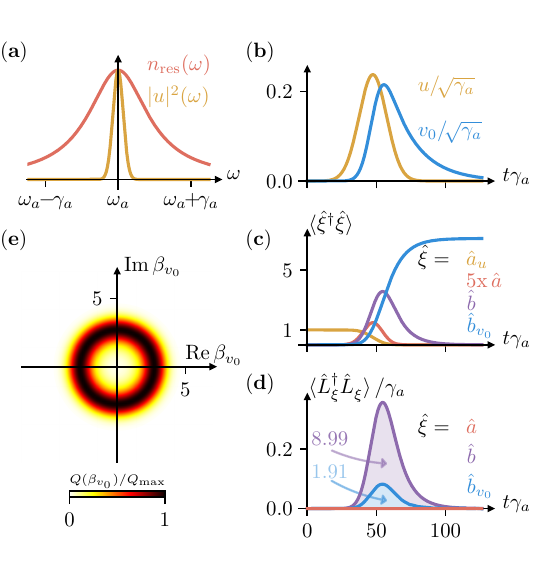}
    \caption{Simulation of the amplification of an incoming single-photon pulse.
    (a) The impinging pulse in the frequency domain, $u(\omega)$, is Gaussian and smaller than the width of the resonance curve $n_\mathrm{res}$ of cavity $\ann$ (occupation of a single cavity biased at $2eV_\mathrm{dc}=\hbar\omega$), so that it can be absorbed. (b) Pulse $u(t)$ in time domain and an output mode $v_0(t)$ of the right transmission [compare Fig.~\ref{fig:fig_model}(a, c)]. They are modeled by auxiliary cavities  with time-dependent loss rates $g_u(t)$ and $g_v(t)$ (not shown) and frequency $\omega_a$ and $\omega_b$. (c) Occupation numbers gained from the simulation show how the input pulse loses its photon to the input cavity $\ann$. The multiplication process driven by the Josephson junction transfers excitations from cavity $\ann$ to cavity $\bnn$, which subsequently leaks photons into the output transmission line. Part of this output is in mode $v_0(t)$ and thus absorbed by the auxiliary cavity modeling that mode. (d) Integrating the rate of lost photons (from $\ann$: 0 lost photons (red), from $\bnn$: $n=9$ lost photons) yields a perfect multiplication to $n$ photons when choosing the optimal Josephson driving strength $E_J$. The explicitly modeled auxiliary cavity $v_0$ from the right transmission line retains $\ex{\hat{n}} = 7.1$ photons [cf. blue line in (c) approaching that constant], but loses $1.91$ photons, which are contained in other, not explicitly simulated modes. (e) The mode $v_0$ is found in a mixture of Fock states as shown by its Husimi-Q function.  \newline
    [Parameters: $n=9$, $ \gamma_b =\sigma_\omega = \gamma_a/10$, $E_J^* = E_{J, \mathrm{opt}}^*$].
    }
    \label{fig:fig_simulation}
\end{figure}

As an introductory example of the simulation and its typical observables for this setting, we consider in Fig.~\ref{fig:fig_simulation} a single-photon Fock state $\ket{1_u}$ in a Gaussian temporal pulse envelope $u(t)$ as incoming quantum pulse of the input transmission line. This itinerant pulse is resonant with cavity $\ann$ [compare Fig.~\ref{fig:fig_simulation}(a, b)], such that it will be completely absorbed. In the output transmission line, we model a single outgoing pulse in a mode $v_0(t)$ specified in Fig.~\ref{fig:fig_simulation}(b). Our results [Fig.~\ref{fig:fig_simulation}(c)] show how the photon leaks from the auxiliary cavity (i.e., the input transmission line) and is absorbed by cavity $\ann$. Subsequently, the transfer process with a multiplication factor $n=9$, driven by the Josephson junction, starts to accumulate photons in $\bnn$. Some of the photons emitted from cavity $\bnn$ into the output transmission line are  absorbed by the auxiliary cavity, indicating that they are emitted into mode $v_0(t)$ of the output transmission line.
Its final state, is characterized by the Husimi-Q function in Fig.~\ref{fig:fig_simulation}(d), corresponds to the quantum state of that mode, which in our case is a diagonal mixture of Fock states $\ket{0},...\ket{n}$. 

In general, the transfer process is not perfect. First, note that a photon loss of cavity $\ann$ to the left transmission line prevents any photon multiplication. Since the Cooper-pair driven multiplication process between the microwave cavities of the \AMP\ is coherent, i.e. a Hamiltonian term, the reverse process is also possible, further increasing the probability of a loss of an $\ann$ excitation into the input transmission line. There is, however, a specific driving strength, $E_{J, \mathrm{opt}}^* = \hbar \sqrt{\gamma_a \gamma_b} \sqrt{n n!} / \alpha_0 \beta_0^n $ \cite{Juha_ICTA_2018}, where one can achieve (near-)perfect conversion [Nominally, it becomes perfect for an incoming single photon in a delta-pulse mode with $\omega_\mathrm{in} = \omega_a$]. For $E_{J, \mathrm{opt}}^*$  all portions of the input pulse that are well within the frequency-dependent bandwidth $T(\omega)$ \cite{Juha_ICTA_2018} of the amplifier are converted. In fact, the conversion probability, $p_\mathrm{conv} = 1 - \int \ex{\hat{L}_a^\dagger\hat{L}_a} dt = \int  \ex{\hat{L}_b^\dagger\hat{L}_b} dt / n$, which can be defined via the integral with respect to time over the lost photons from either $\ann$ or $\bnn$, cf. Fig.~\ref{fig:fig_simulation}(d), can be expressed as $p_\mathrm{conv} = \int T(\omega) |u(\omega)|^2 d\omega$.
Notably, the optimal driving strength can also be intuitively derived by matching the rates $\gamma_a$ of the absorption of the input photon into cavity $\ann$ to a golden-rule rate $\Gamma_\mathrm{gr} \propto |\bra{0_a, n_b} \hat{H}_{J,\mathrm{RWA}} \ket{1_a, 0_b}|^2 / n\gamma_b$. The golden rule rate, which is formally valid only in the perturbative regime, describes the transition probability of the multiplication process followed by the subsequent loss of the first excitation 
from the lifetime-broadened level $\ket{n_b}$. Matching this first photon loss process is sufficient for a perfect conversion, since the loss of the first photon from cavity $\bnn$ effectively brings the coherent dynamics to an abrupt end, as the Josephson drive is not resonant anymore.

\section{Photon Detection in Single-Stage Amplifier}

In fact, a simple optimization of the conversion probability is not sufficient for an eventual heterodyne detection of microwave photons. Of more importance than the total number of photons in the output, is a high occupation of a single mode: as apparent from Fig.~\ref{fig:fig_simulation}(e), the quantum state of the highly-occupied simulated mode can be quite clearly distinguished from a vacuum mode by comparing the Husimi-Q functions. Before analyzing the full photon detection schemes in a heterodyne measurement setup, we therefore first start to investigate parameter regimes of the \AMP\ possessing different output modes.

\subsection{Regimes of Output Modes}
\label{sec::OutputModes}
  In general, the photons leaking from cavity b are distributed among several (eigen-) modes in the output transmission line. An eigenmode decomposition can be found by diagonalizing the first-order two-time correlation function \cite{Treps2020_REV, Devoret2010_REV, Kiilerich2019, Kiilerich2020},
  \begin{equation}
      G^{(1)}(t_1, t_2) = \ex{\bdag_\mathrm{out}(t_2) \bnn_\mathrm{out}(t_1)} = \sum_k n_k v_k^*(t_2) v_k(t_1) \;,
  \end{equation}
  which yields the mean occupation $n_k = \ex{\bdag_{v_k} \bnn_{v_k}}$ of the k-th eigenmode $\bnn_{v_k} = \int_0^\infty \bnn_\mathrm{out}(t) v_k(t) dt$ in the output transmission line.

\begin{figure*}[t]
    \centering
    \includegraphics[width=\linewidth]{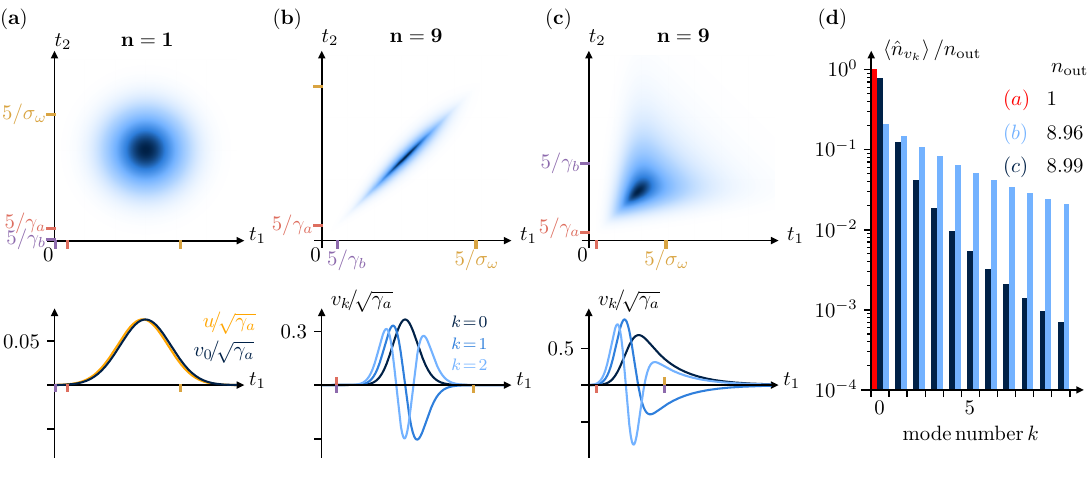}
    \caption{
    Two-time first order correlation functions of the output field, $G^{(1)} = \ex{\bdag_\mathrm{out}(t_2) \bnn_\mathrm{out}(t_1)}$ with eigenmodes and their occupations $\ex{\hat{n}_k} = \ex{\bdag_k \bnn_k}$. In all cases, we simulate a single input mode centered around $\omega_a$ in a Gaussian envelope with $\sigma_\omega = \gamma_a/10$ in the quantum state $\ket{1}_\mathrm{in}$ impinging on and being resonantly absorbed by cavity $\ann$, and $E_J^* = E_{J, \mathrm{opt}}^*$, for optimal conversion to $n$ output photons on resonance.  
    (a) Linear conversion, $n=1$ and $\gamma_b = 10 \gamma_a$, results in one output mode  with occupation $\ex{\bdag_0 \bnn_0} = 1$, cf. (d). 
    (b) For nonlinear multiplication $n=9$ and $\gamma_b = \gamma_a$, the duration of the outgoing pulse (the extent of $G^{(1)}$ along the diagonal) is much longer than the inverse instantaneous bandwidth of the amplifier (off-diagonal extent). $G^{(1)}$ becomes elliptical and possesses multiple eigenmodes with similar occupations. 
    (c) For $\gamma_a = 10 \gamma_b$, the pulse duration of the output pulse is comparable to the inverse bandwidth of the instantaneous spectrum, yielding a triangular $G^{(1)}$ diagonalizable to very few dominant eigenmodes, see (d). 
    Ticks mark different time scales that help to characterize 
    $G^{(1)}$.
    }
    \label{fig:fig_1stage_modes_regimes}
\end{figure*}

It will turn out to be advantageous for detection to dominantly occupy a single output mode. To understand how to achieve a single-mode output, we can preliminarily consider the simplest scenario of this type, which can be realized in our amplifier for a multiplication factor $n=1$.
For that case, the bilinear Hamiltonian \eqref{eq::1stage_Hamiltonian_RWA}  can be solved analytically, and a scattering matrix connects input and output modes at frequencies linked by the biasing condition. 
Assuming vacuum input into the output transmission line and a single-photon input into the input transmission line, we find a single output mode, which, at optimally chosen Josephson driving strength, carries the full occupation $n_0 = 1$ [see Fig.~\ref{fig:fig_1stage_modes_regimes}(a)]. Approaching the limit of continuous-wave input (i.e., a very  sharp pulse shape in Fourier space with a frequency spread $\sigma_\omega$ smaller than all other rates), the extent of the coherence function in the diagonal direction $t_+ = (t_1 + t_2) / 2$ grows to a correspondingly long time $T_\mathrm{out} \sim 1/\sigma_\omega$. 
An understanding of other directions in the $G^{(1)}(t_1, t_2)$ map can be gained by considering vertical and horizontal cuts starting at the diagonal as the Fourier transform of the instantaneous spectrum at time $t_1=t_2$. 
In Fig.~\ref{fig:fig_1stage_modes_regimes}(a), the long extent of the correlation function in these directions, $T_\mathrm{spec}\sim 1/\sigma_\omega$, and the corresponding circular shape of $G^{(1)}(t_1,t_2)$ thus reflects the fact that the frequency of the outgoing photon for $n=1$ is fixed by the resonance condition $\omega_b  = \omega_\mathrm{dc} + \omega_\mathrm{in}$, up to a broadening arising from the input photon bandwidth $\sigma_\omega$.
Crucially, just as in a linearly driven damped harmonic oscillator, the instantaneous spectrum is not lifetime broadened by $\gamma_b \gg \sigma_\omega$. Such broadening would result in a $G^{(1)}$ more strongly confined along the diagonal.

Such a confinement indeed occurs in the nontrivial cases of photon amplification with $n>1$, see e.g.~Fig.~\ref{fig:fig_1stage_modes_regimes}(b). The resonance condition $\omega_\mathrm{in} = \omega_\mathrm{dc} + n \omega_b$ now only requires the sum of the frequencies of all $n$ output photons to match the frequency distribution of the input photon. The spectral distribution of output photons is now lifetime-broadened by $\gamma_b$ adding to the spectral broadening of the input $\sigma_\omega$, so that, in the time domain, $G^{(1)}$ can become substantially sharper in the off-diagonal directions.  

Choosing a rather large (but still realizable) multiplication factor of $n=9$, we will now discuss parameter regimes with different relations between $T_\mathrm{out}$ and $T_\mathrm{spec}$ of the $G^{(1)}$ function [Fig.~\ref{fig:fig_1stage_modes_regimes}(b)-(d)] and the resulting complex multimode structure.

We start by considering a resonant Gaussian single-photon input pulse (i.e., it contains frequencies $\omega_\mathrm{in} \sim \omega_a \pm \sigma_\omega$ where $\sigma_\omega \ll \gamma_a, \gamma_b$ and choose the optimal Josephson driving energy (fixed by the rate matching argument to achieve the best possible photon multiplication). We are then left with one parameter to vary, namely the ratio of the loss rates $\gamma_a/\gamma_b$ \footnote{It is easy to see, that the zero-point fluctuations $\alpha_0,\, \beta_0$ merely enter as trivial prefactors of the Josephson driving energy, since for a single-photon input only the transition matrix element between two states $(n_a=1,\,n_b=0) \Leftrightarrow (n_a=0,\,n_b=n)$ matters.}. 

At $\gamma_b = \gamma_a$, Fig.~\ref{fig:fig_1stage_modes_regimes}(b), we find an elongated elliptic-like shape of $G^{(1)}$. In this regime, the frequency distribution of the output photons is spectrally broadened by $\gamma_b \gg \sigma_\omega$, resulting in a short time extent away from the diagonal, $T_\mathrm{spec} \sim 1/\gamma_b$, as compared to the long duration of the output pulse along the diagonal, $T_{out} \sim 1/\sigma_\omega$.  Diagonalization of such elliptic shapes with principal axes of strongly different lengths  \cite{RasmussenGaussianMachineLearning} ultimately yields a large number of eigenmodes with similar mean occupations, cf. Fig.~\ref{fig:fig_1stage_modes_regimes}(b). 

Following what we learned from the $n=1$ case, we turn to the regime where $\gamma_b \ll \gamma_a$ [Fig.~\ref{fig:fig_1stage_modes_regimes}(c)], where we expect the reduced $\gamma_b$ to increase $T_\mathrm{spec}$ and thus a $G^{(1)}$ with comparable extent along and off the diagonal. Indeed, the $G^{(1)}$ function is of triangular shape with two similar timescales and can be diagonalized into very few eigenmodes with one strongly occupied main eigenmode carrying $\sim 75\,\%$ of the emitted photons. Its quantum state can thus be easily distinguished from vacuum [cf. Fig.~\ref{fig:fig_simulation}(e)]. It is this regime, where $G^{(1)}$ is as 'symmetric' as possible and yields only one highly occupied mode, that will prove well-suited for the eventual detection of microwave photons.  

\subsection{Heterodyne Detection with a highly occupied eigenmode}
\label{sec::Detection1stage_Scheme1}

 In a typical experiment \cite{Juha_ICTA_2018, Martel_ICTA_2025, Albert_ICTA_2024}, a signal from the transmission line will first be amplified by a linear phase-preserving amplifier with gain $G$ before recording both quadratures with a classical device. For a quantum-limited amplifier, the noise mode $\hat{r}$ can additionally be assumed to be in a vacuum state, so that the minimal noise required by quantum mechanics is added. For $G\gg1$, the signal after amplification, $\hat{J}^{(b)}(t) = \sqrt{G} \bnn_\mathrm{out}(t) + \sqrt{G-1} \hat{r}^\dagger$ can be treated as a classical variable in the subsequent signal processing \cite{daSilva2010}. The sketched experimental scenario corresponds to a quantum-optical heterodyne detection scheme, which can theoretically be described with a stochastic Schr\"odinger equation \cite{WisemanQuantumMeasurement}. From the time-evolution (compare Appendix \ref{sec::app_cascadedME}), we obtain single trajectories $J_k^{(b)}(t) =  \ex{\bnn_\mathrm{out}}_k + \xi_k(t)$. The stochastic white noise term $\xi_k(t)$ with mean $\mathrm{E}\left[\xi_k(t)\right]=0$ is delta-correlated $\mathrm{E}\left[\xi_k(t+\tau)\xi_k(t)\right] \propto \delta(\tau)$ and simulates vacuum noise. 
The simulated $J_k^{(b)}(t)$ for a single trajectory then corresponds to the classical record of $J^{(b)}(t)/\sqrt{G}$ measured in a single run in its signal-to-noise ratio and all other statistical properties.

\begin{figure}[htb]
    \centering
    \includegraphics[width=\columnwidth]{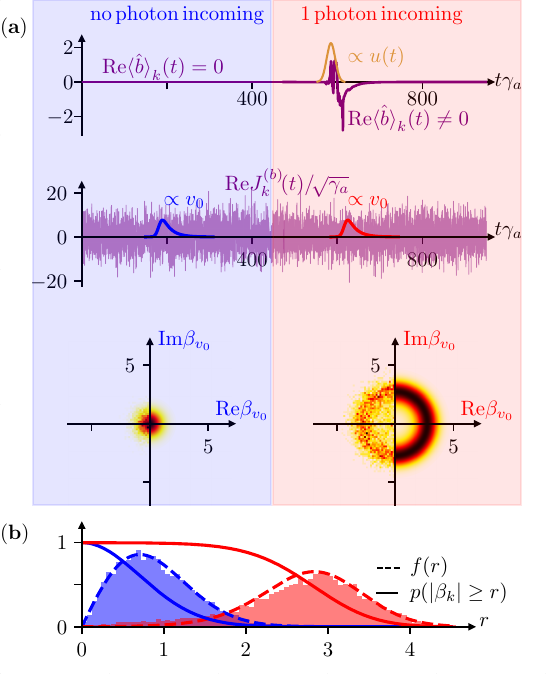}
    \caption{(a) Weak measurement of the output from cavity $\bnn$ by heterodyne detection. A single trajectory of the numerical simulation yields $\ex{\bnn} = 0$, without incoming photon, and $\ex{\bnn} \neq 0$ during the photon multiplication process of the \AMP\ when a single photon (Gaussian input mode in Fock state $\ket{1}_\mathrm{in}$, orange curve) impinges on cavity $\ann$. Experimentally observable, however, is only the measured signal $J_\beta$ dominated by noise, where information about photon arrivals seems lost. Integrating the measured signal with a mode envelope $v_0$ results in a single complex number, which samples the Husimi-Q function of $v_0$ at the specific integration time. (b) The probability density  $f(r)$ for the radial component, $|\beta_k| =r$, clearly discriminates cases without an incoming photon (blue histogram and dashed theory curve) from the highly occupied output mode of the \AMP\ after a photon arrival (red histogram and dashed theory curve). Defining a photon detection event, as  $|\beta_k|>r$, for any chosen threshold $r$, the integrals $p(|\beta_k|\geq r) =\int_{r}^\infty d|\beta_k| \, f(|\beta_k| \,| 1_\mathrm{in})$ (solid red line), $1-p(|\beta_k|\geq r)$, and $\int_{r}^\infty d|\beta_k| \,f( |\beta_k| \, | 0_\mathrm{in})$ (solid blue line) are the probability for correct, false negative, and false positive detection. 
    [Parameters: $N_\mathrm{traj} = 10^4$ trajectories simulated,  
    $\sigma_\omega = \gamma_b = \gamma_a / 10$, $n=9$, $E_J^* = E_{J, \mathrm{opt}}^*$].
    }
    \label{fig:fig_1stage_scheme1_1}
\end{figure}

If vacuum is sent through the input transmission line to the \AMP, cavity $\bnn$ will remain in its ground state with $ \ex{\bnn} = 0$. If, on the other hand, a single-photon pulse is sent, the information acquired by the weak heterodyne measurement induces a feedback on the system's time evolution, such that cavity $\bnn$ will explore its phase-space and $\ex{\bnn} \neq 0$ during the response time of the \AMP. Note that the modification of the system's time evolution by the quadrature measurement merely reflects our incrementally increasing knowledge about the system's state.  Experimentally, circulators prevent any physical signal backflow from the detection apparatus to the system. Only in a numerical simulation is the quantity  $\ex{\bnn}_k$ known for any single noise realization. In an experiment, the only accessible observable is $J_k^{(b)}$, which is dominated by noise, see Fig.~\ref{fig:fig_1stage_scheme1_1}(a). 
It is thus a crucial task to develop a reliable scheme for detecting the small signal created by an incoming photon in each single noise-dominated trajectory $J_k^{(b)}$. This will be provided in the sequel.

Assuming we knew that for a certain trajectory $J_k^{(b)}(t)$ a single-photon pulse impinged on the system at a certain arrival time, we could integrate the measured signal with the dominant output eigenmode corresponding to that arrival time, $\beta_k = \int_0^\infty v_0(t) J_k^{(b)}(t) dt$, to get one complex-valued number. If the same experiment is repeated multiple times, the mode-matched values $\beta_k$ can be plotted in a histogram [see Fig.~\ref{fig:fig_1stage_scheme1_1}(a)]. This histogram samples the Husimi-Q function of the quantum state of the corresponding mode $\bnn_{v_0} = \int_0^\infty v_0(t) \bnn_\mathrm{out}(t) dt$ \cite{Kim1997, daSilva2010,  MilburnCarmichael2010, Eichler2011}. When $v_0$ represents a highly occupied mode, the probability $p(\beta_k |\,1_\mathrm{in})$ to find $|\beta_k|$ at larger absolute values is greatly enhanced for a single-photon input compared to the corresponding probability $p(\beta_k |\,0_\mathrm{in})$ for a vacuum state [Fig.~\ref{fig:fig_1stage_scheme1_1}(b)]. By defining a click threshold, $r_0$, a simple decision process identifies photons in a trajectory $k$ if $\beta_k>r_0$. For this scheme, it is therefore crucial to fabricate an \AMP\ device which exhibits one highly occupied output mode, cf. Sec. \ref{sec::OutputModes}. Then $r_0$ can be defined large enough to suppress dark counts, $p_\mathrm{dark} =p(|\beta_k| >r_0 |\, 0_\mathrm{in})\ll 1$, while keeping the photon detection probability $p_\mathrm{click} = p(\beta_k > r_0 |\, 1_\mathrm{in})$ sufficiently large. 
 
\begin{figure}[t]
    \centering
    \includegraphics[width=\columnwidth]{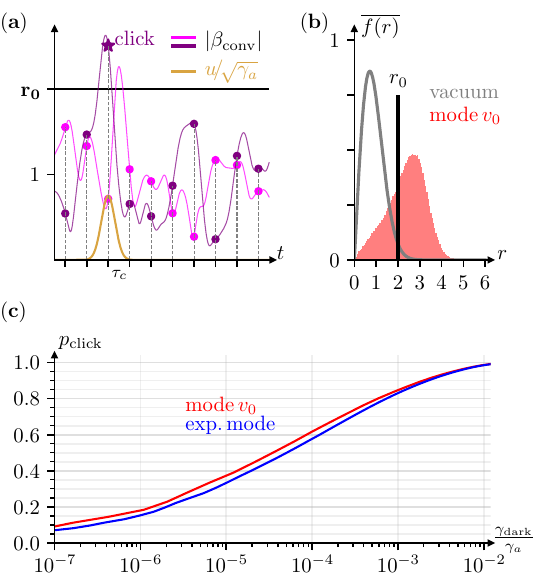}
    \caption{
    (a) Convolution $\beta_\mathrm{conv}$ of two recorded trajectories with a single mode $v_0$, for a simulation with an input photon in a Gaussian mode $u(t)$.
    The convolution induces correlations within a correlation time $\tau_\mathrm{c}$.
    Measuring stroboscopically once within $\tau_\mathrm{c}$ (circles) avoids induced two-time correlations and reduces the effective number of measurement points. A photon detection event at time $t$ is defined when $|\beta_\mathrm{conv}|>R_0$.
    (b) The resulting probability distributions, where averaging the instances of stroboscopic measurement over an interval of length $\tau_\mathrm{c}$ for an unbiased comparison to experiment results in a slight deterioration  compared to Fig.~\ref{fig:fig_1stage_scheme1_1}(b). 
    (c) The resulting performance curve showing detection probability versus the dark-count rate for convolution with the dominant eigenmode $v_0$ and an exponential mode $v_\mathrm{exp}(t) \propto \exp[- \gamma_bt/2]$. The comparison with an exponential mode with comparable timescales demonstrates that a small mismatch between detection mode and  eigenmode (or similarly between assumed and actual input mode) only slightly degrades the performance.
    [Parameters see Fig.~\ref{fig:fig_1stage_scheme1_1}].
    }
    \label{fig:fig_1stage_scheme1_2}
\end{figure}

In a real experimental setup, which tries to identify the arrival of single microwave photons at unknown arrival times within a long measurement trajectory, the obvious idea is now to replace the integration with the mode $v_0$ by a convolution, $\beta_k \rightarrow \beta_k(t) = \int_0^\tau v_0(t-\tau) J_k^{(b)}(\tau) d\tau$. This will introduce correlations between the values $\beta_k(t)$ within a correlation time $\tau_\mathrm{c}$ set by the mode shape $v_0$ (beyond any correlations that may be present within the dynamics of $\hat{b}_\textrm{out}$). One simple adequate method to get roughly uncorrelated discrete measurement points and thereby avoid double-counting a single photon while also keeping the dark count rate low is to stroboscopically take one measurement point within each time interval $\tau_\mathrm{c}$, cf. Fig.~\ref{fig:fig_1stage_scheme1_2}(a). 
  
The stroboscopic scheme focuses on detecting single photons and more elaborate schemes will be needed to detect two (or more) nearly coincidental photons. To avoid such a scenario of overlapping modes,  the mean rate of incoming photons in an experiment should be kept sufficiently small, $\gamma_\mathrm{ph} < \tau_\mathrm{c}^{-1}$. 
Employing the presented scheme, each chosen threshold value will result in a pair of numbers: t
The photon detection probability and the associated dark count rate, which can be combined in a single performance curve, see Fig.~\ref{fig:fig_1stage_scheme1_2}(c). Here, we report, for instance,  an achieved photon detection probability of $p_\mathrm{click} \approx 0.85$ for a dark count rate $\gamma_\mathrm{dark} = 5 \cdot 10^{-4} \gamma_a$, where filtering with the highest occupied eigenmode induced a correlation time $\tau_\mathrm{c}\approx 30/\gamma_a$.
For an experiment with an average photon rate of $\gamma_\mathrm{ph} = \gamma_a/150 \approx 1/(5 \tau_\mathrm{c})$, this means that within a time $\Delta t = 15 \cdot 10^3 / \gamma_a$ our scheme will correctly detect $78.3$ while producing $7.5$ dark counts, where we expect an average of $100$ incoming photons. We explain specific steps how these numbers can be improved further below in Sec.~\ref{sec:two-stage}.

\subsection{Detection in the multi-mode regime}
The detection scheme of Sec.~\ref{sec::Detection1stage_Scheme1} crucially relies on an output, where one eigenmode has a high occupation. That allows us to clearly distinguish the radial distribution of the Husimi-Q function of that mode from the vacuum one.  
Turning now to a parameter regime, where there are \emph{many} eigenmodes, each with low occupation,  cf. Fig.~\ref{fig:fig_1stage_modes_regimes}(b), the question arises: How to detect the signature of an impinging photon? 

\begin{figure}[t]
    \centering
    \includegraphics[width=\columnwidth]{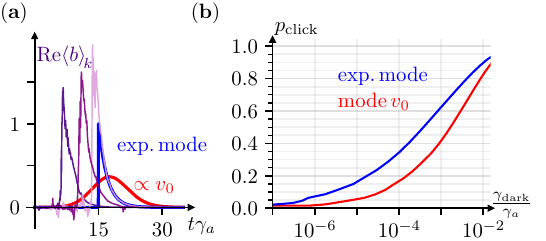}
    \caption{
    (a) The  expectation value $\ex{\bnn}$ for individual trajectories (shades of purple), which (while not measurable) is available from numerical simulations, can be used to find a suitable detection mode [here, a simple exponential mode $v_\mathrm{exp}(t) = \sqrt{\gamma_b} \theta(t-t_c)\exp[- \gamma_b(t-t_c)/2]$ (blue)]. In a regime of multi-mode output, that mode performs better than the highest-occupied eigenmode $v_0$ (red) as seen by comparing the curves of
    detection probability vs. dark count rate in (b).\newline
    [Parameters: $N_\mathrm{traj} = 10^4$ trajectories simulated,  
    $\sigma_\omega = \gamma_a / 10$, $\gamma_b = \gamma_a$, $n=9$, $E_J^* = E_{J, \mathrm{opt}}^*$].
    }
    \label{fig:fig_1stage_scheme2}
\end{figure}

Technically, this means to retrieve a finite expectation value $\sqrt{\gamma_b} \ex{\bnn}$ in the background of a noisy measurement trajectory $J_k^{(b)}(t) =  \sqrt{\gamma_b}\ex{\bnn}_k + \xi_k(t)$. To identify a suitable photon-detection mode, the expectation values $\ex{\bnn}$ of numerically simulated trajectories, see Fig.~\ref{fig:fig_1stage_scheme2}(a), can be used. Their shape can be approximated by an exponential function, $f(t) \propto \exp[-\gamma_b (t-t_0)/2]$ (blue curve) for times larger than the arrival time, $t \ge t_0$.
If the recorded measurement is filtered with $f(t)$, we can thus expect that the filtered measurement peaks at the arrival times of photons.  The fact that the highest-occupied eigenmode [with a Gaussian shape, cf. red curve in Fig.~\ref{fig:fig_1stage_scheme2}(a)] does not match the typical shape of $\ex{\bnn}$, and therefore yields an imperfect overlap, makes filtering with this eigenmode unfavorable in this parameter regime. Fig.~\ref{fig:fig_1stage_scheme2}(b) confirms this reasoning: Filtering the measurement with an exponential mode yields noticeably better probabilities for photon detection. Nonetheless, compared to the example given in Sec. \ref{sec::Detection1stage_Scheme1}, we will now only detect $53.3$ of $100$ incoming photons (given the same number of  $7.5$ dark counts for a photon rate of $\gamma_\mathrm{ph} = \gamma_a / 150$).

To draw preliminary conclusions from the scenarios presented above, it clearly is beneficial to design a detector in such manner that it produces \emph{one highly-occupied output eigenmode}. Moreover, the detection probability is obviously enhanced if the multiplication factor $n$ is increased. 
Experimentally, multiplication factors are probably limited to $n \lesssim 5-10$, as either Josephson energy $E_J$, or zero-point fluctuations, $\alpha_0,\beta_0$, have to be increased to achieve a sufficiently large transition matrix element for the multiplication process, which brings about such complications as the breakdown of the rotating wave approximation and the appearance of undesired competing resonances. One way to overcome such experimental limitations is a cascaded version of the \AMP, where the multiplication process happens in two stages as will be discussed in the remainder of this paper.

\begin{figure*}[t]
    \centering
    \includegraphics[width=\linewidth]{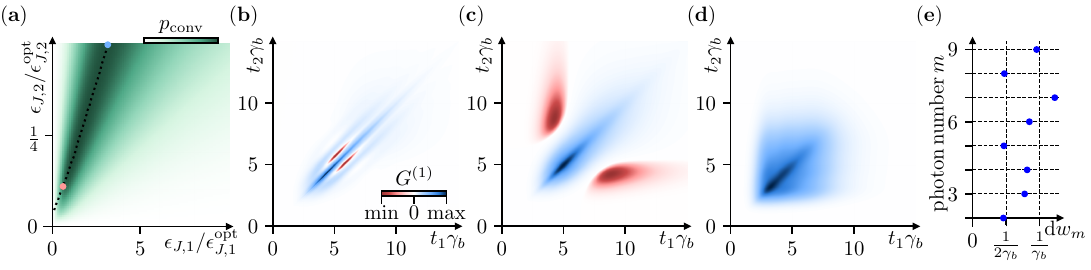}
    \caption{
    (a) Probability of photon conversion for the two-stage \AMP\ for multiplication factors $n_1=n_2 =3$ (max. $9$ output photons). Given $\epsilon_{J1}$, best conversion is achieved for $\epsilon_{J2}$ along the black dotted line.
    (b-d) Two-time first-order coherence functions $G^{(1)}$ yield a coherence pattern due to Rabi-oscillations in (b) [parameters marked with blue dot in (a)] with multiple weakly occupied eigenmodes (max. mean photon occupation is $\ex{\bdag_{v_0} \bnn_{v_0}}=1.52$ photons). 
    (c) Reducing $\epsilon_{J2}$ [red dot in (a)] improves the occupation $\ex{\bdag_{v_0} \bnn_{v_0}}=2.89$ of the largest-occupied eigenmode, which is however insufficient due to poor overall photon conversion. 
    (d) Zero-point fluctuations $\kappa_0=0.2$ and $\beta_0 \approx 1.66$ can suppress internal \AMP\ processes to obtain a few-mode regime with one dominant eigenmode, $\ex{\bdag_{v_0} \bnn_{v_0}}=6.42$.
    (e) Mean waiting times $dw_m=w_{m, m-1}$ of the $m$-th photon emitted from cavity $\bnn$ for parameters in (d), obtained by a quantum jump approach (explanation given in main text). \newline
    [Parameters: $\sigma_\omega = \gamma_b=\gamma_a/100$ 
    (a-c): $\kappa_0, \beta_0 \ll 1 $ limit; (d): $\epsilon_{J,1}/\epsilon_{J,1}^\textrm{opt}=0.9,\,\epsilon_{J,2}/\epsilon_{J,2}^\textrm{opt}=0.1$; color scales in (b)-(d) stretched, where [min, max] is (b) [-0.224,\, 2.233] (c) [-0.031,\,1.852] (d) [0,\, 1.388] in units of $\gamma_b$].
    }
    \label{fig:fig_2stage_modes}
\end{figure*}

\section{Photon Detection in Two-Stage Amplifier}
\label{sec:two-stage}

The idea of photon multiplication in two stages is straightforward. The two-stage \AMP\ [compare Appendix \ref{sec::app_2stage_RWA}, Fig.~\ref{fig:app_2stage}(a)] incorporates a third central cavity $\cnn$ without input or output lines, so that $\gamma_c \approx 0$, and with zero-point fluctuations $\kappa_0$. 
Two dc-driven processes through two Josephson junctions now drive multiplication in two separate stages, if resonance conditions are met: $2eV_{\mathrm{dc},1} = \hbar (n_1 \omega_c - \omega_a)$, transfers one impinging photon from cavity $\ann$ to $n_1$ photons in cavity $\cnn$, and $2eV_{\mathrm{dc},2} = \hbar (n_2 \omega_b - \omega_c)$, transfers each photon from cavity $\cnn$ to $n_2$ photons in cavity $\bnn$. While the basic working scheme has been easily generalized to two (or even more) stages, the internal dynamics of a multi-stage amplifier becomes more complex.  
For the single-stage \AMP\ and single-photon input, there is a single relevant coherent process, namely $\ket{1}_a\ket{0}_b\leftrightarrow \ket{0}_a \ket{n}_b$. Specifically, after the first photon has leaked from cavity $\bnn$ to the output transmission line, coherent back-transfer to cavity $\ann$ (and subsequent loss to the left output) becomes impossible. 
For the two-stage \AMP\ only the first multiplication stage  behaves similarly (coupling only the two states $\ket{1}_a\ket{0}_c\leftrightarrow \ket{0}_a \ket{n_1}_c$.) 

The complete transfer of  $n_1$ photons in cavity $\cnn$  to cavity $\bnn$ in the second amplification stage involves transitions other than the one between $\ket{n_1}_c\ket{0}_b \leftrightarrow \ket{n_1-1}_c \ket{n_2}_b$. To be specific,  let us consider an example where one photon from cavity $\cnn$ has been multiplied to $n_2$ photons in cavity $\bnn$ arriving at a state $\ket{n_1-1}_c \ket{n_2}_b$. Cavity  $\bnn$ may now lose one or more photons, so that the consecutive multiplication process may involve the transitions $\ket{n_1-1}_c \ket{m}_b \leftrightarrow \ket{n_1-2}_c \ket{m + n_2}_b$ with $m=0...,n_2$. 
In fact, many of these coherent forward and backward photon-transfer processes in the second multiplication stage compete with the progressive loss of photons from the output cavity $\bnn$ until all but $n_2-1$ photons have leaked out.

The nonlinearity of the Josephson junction and the values of the zero-point fluctuations, $\beta_0,\kappa_0$, affect the Hamiltonian matrix elements of these transitions of the second stage.
To understand how, consider the two parts of the driving Hamiltonian in the lab frame $ \hat{H}_{J} =  \hat{H}_{J1} +  \hat{H}_{J2}$, 
\begin{equation}
 \begin{split}
    \hat{H}_{J1} &= -E_{J1} \cos\left[\frac{2e}{\hbar}V_{\mathrm{dc},1}t + \alpha_0 (\adag + \ann) + \kappa_0 (\cdag + \cnn)\right] \\
    \hat{H}_{J2} &= -E_{J2} \cos\left[\frac{2e}{\hbar}V_{\mathrm{dc},2}t + \kappa_0 (\cdag + \cnn) + \beta_0 (\bdag + \bnn)\right] \, .    
 \end{split}
 \end{equation}
After moving to a rotating frame (using the resonance condition), we arrive for the second stage at $ \hat{H}_{\mathrm{RWA}, J2} = \hat{h}_2^\dag + \hat{h}_2$ with 
 \begin{equation}
 \label{eq::2stage_Hamiltonian_RWA_stage2}
  \hat{h}_2 =  \frac{\epsilon_{J,2}}{2 n_2!} :\cnn \bdag[n_2] \frac{J_1\left(2\kappa_0\sqrt{\cdag \cnn}\right)}{\kappa_0 \sqrt{\cdag \cnn}} \frac{J_{n_2}\left(2\beta_0\sqrt{\bdag \bnn}\right)\cdot n_2!}{\left(\beta_0 \sqrt{\bdag \bnn}\right)^{n_2}}:  \, 
 \end{equation}
which now contains normal-ordered Bessel functions (cf. Appendix \ref{sec::app_2stage_RWA}), while the first-stage term $\hat{H}_{J1}$ turns into  Eq.\,\eqref{eq::1stage_Hamiltonian_RWA} with $(\beta_0,\hat{b}^{(\dagger)})$ replaced by $(\kappa_0,\hat{c}^{(\dagger)})$. The Josephson energies are parametrized as $\epsilon_{J,1} =  E_{J,1}^*\alpha_0 \kappa_0^{n_1}$ and $\epsilon_{J,2} = E_{J,2}^*\kappa_0 \beta_0^{n_2}$, where we will also denote values $\epsilon_{J,1}^\mathrm{opt} = \hbar \sqrt{\gamma_a \gamma_b} \sqrt{n_1 \cdot n_1!}$ and analogously $\epsilon_{J,2}^\mathrm{opt} = \hbar \sqrt{\gamma_a \gamma_b} \sqrt{n_2 \cdot n_2!}$, which would give optimal conversion in a single-stage \AMP. 

Remarkably, for a single resonant impinging photon, perfect photon multiplication to $n_1 \cdot n_2$ outgoing photons is still possible. Indeed, for the linear case, $n_1=n_2=1$, one finds the condition \cite{Juha_ICTA_2018},
\begin{equation}
 \label{eq:rate_matching_pconv_2stage}
    \sqrt{\gamma_a} \frac{\epsilon_{J,2}}{\epsilon_{J,2}^\mathrm{opt}} = \sqrt{\gamma_b} \frac{\epsilon_{J,1}}{\epsilon_{J,1}^\mathrm{opt}} \, .
 \end{equation}
This expression also gives the correct scaling for the case of general multiplication factors and small zero-point fluctuations, see Fig.~\ref{fig:fig_2stage_modes}(a)  where $n_1=n_2=3$ and $\gamma_a = 100\gamma_b$.

However, contrary to the single-stage case, where the rate-matching expression holds throughout, the dynamics of the two-stage \AMP\ for generic multiplication factors is more complex, and the strictly linear relation \eqref{eq:rate_matching_pconv_2stage} is replaced by the line of optimal conversion in Fig.~\ref{fig:fig_2stage_modes}(a). 
In general, both the zero-point fluctuations, $\beta_0, \kappa_0$, via their effect on the transition matrix elements between higher Fock-states, and the two Josephson energies $\epsilon_{J,1/2}$ affect the conversion probability.  In consequence, we will have to refrain from a complete systematic exploration of the full parameter space in our goal of finding a regime of a highly occupied eigenmode of the output field, which can be used for detecting photons analogously to the scheme presented in Sec. \ref{sec::Detection1stage_Scheme1}. 
Instead, drawing on the lessons from the single-stage case and our understanding of the impact of nonlinearities and zero-point fluctuations, we will show how to locate a favorable regime with a highly occupied output mode in Sec. \ref{sec::output_modes_2stage}, before analyzing achievable detection efficiencies in Sec. \ref{sec::sec_detection_2stage}.

\subsection{Output Modes for the two-stage \AMP}
\label{sec::output_modes_2stage}

We first analyze the output modes in the limit of zero-point fluctuations,  $\kappa_0, \beta_0 \ll 1$, where the Bessel functions in \eqref{eq::2stage_Hamiltonian_RWA_stage2} and thus any nontrivial dependence of the results on $\kappa_0$ and $\beta_0$ can be neglected. 
Following the lessons learned in Sec.~\ref{sec::OutputModes}, we first choose $\gamma_b = \sigma_\omega \ll \gamma_a$
and pick a rather large value $\epsilon_{J,2} = \epsilon_{J,2}^\mathrm{opt}/2$,
on the line of optimal overall conversion in Fig.~\ref{fig:fig_2stage_modes}(a) to analyze the $G^{(1)}$-function in Fig.~\ref{fig:fig_2stage_modes}(b). 

The coherence function reveals a multi-mode structure with pronounced coherence fringes. The number of photons leaked to the output line, $\int_0^t dt' G^{(1)}(t',t')$, indicates that the time interval, when pronounced coherence fringes occur, starts once about three photons have leaked and lasts until less than three photons remain in the \AMP\ cavities. Simulating quantum jump trajectories furthermore confirms that after three photons have leaked the system is preferentially found with all $6$ remaining photons in cavity $\bnn$. Taking into account the various timescales one can identify in the system, this suggests an evolution, where a coherent dynamic between a highly occupied cavity $\bnn$ and $\cnn$ sets in only after several multiplication processes from $\cnn$ to $\bnn$ have taken place and some photons have leaked, and that this dynamics stops once only a few photons are left. 

Apparently, the interplay of different processes with drastically different timescales leads to a multi-mode emission, which (as learned above) is not conducive to detection. 
To avoid the scenario described above, we hence try to reduce the rate of photon accumulation in $\bnn$.
Achieving this by quenching the Josephson coupling strength by a factor of ten to $\epsilon_{J,2}= \epsilon_{J,2}^\mathrm{opt}/20$ should yield similar rates for the conversion to cavity $\bnn$ and the decay rate from it, thus avoiding the piling up of a large occupation in $\bnn$. However, for such a small $\epsilon_{J,2}$, the rate matching expression of Eq. \eqref{eq:rate_matching_pconv_2stage} fails and we find photon reflection into the input transmission line and poor photon conversion, cf. the regime for small $\epsilon_{J,2}$ in Fig.~\ref{fig:fig_2stage_modes}(a).
Although now the $G^{(1)}$ function shows fewer fringes, see Fig.~\ref{fig:fig_2stage_modes}(c), and the dominant eigenmode carries more photons than before ($2.9$ of $7.7$ outgoing photons), the distinguishability from a vacuum state is still too weak for a satisfying photon detection.

A second option to avoid large photon accumulation in $\bnn$ involves finite zero-point fluctuations. To delay the second photon multiplication process from $\cnn$ to $\bnn$, until $\bnn$ has emptied, we can fix $\beta_0^2\approx 1.66$ to a root of the Laguerre polynomial appearing in the nonlinear corrections of the Hamiltonian matrix elements $\propto L_2^{(3)}(\beta_0^2)$, corresponding to the transfer $(m_a, m_b=2, m_c) \leftrightarrow (m_a, 5, m_c-1)$, 
see Eqs.~(\ref{eq::2stage_laguerre_1}), (\ref{eq::2stage_laguerre_2}). 
The waiting-time distributions of the leaking photons in Fig.~\ref{fig:fig_2stage_modes}(f), which can be obtained by simulations with a quantum jump approach, nicely confirm this picture: As we expect, the photons number two, five and eight leak out after a mean waiting time of $1/(2\gamma_b)$, while the photons number three, six and nine leak after $\sim 1/\gamma_b$, where these two different waiting times match the average decay time of the states $\ket{2}_b$ and $\ket{1}_b$, respectively. The waiting times of the other photons (number four and seven) are associated with the conversion of a cavity $\cnn$-photon into cavity $\bnn$-photons and one subsequent decay, for which similar timescales $\gamma_b^{-1}$ are observed.
In this regime, we find the desired $G^{(1)}$-function, which is dominated by a highly occupied single mode, cf. Fig.~\ref{fig:fig_2stage_modes}(d) with mean occupation of $6.42$ out of $8.87$ transmitted photons.

\subsection{Heterodyne Detection for the two-stage \AMP}
\label{sec::sec_detection_2stage}

\begin{figure}[b]
    \centering
    \includegraphics[width=\linewidth]{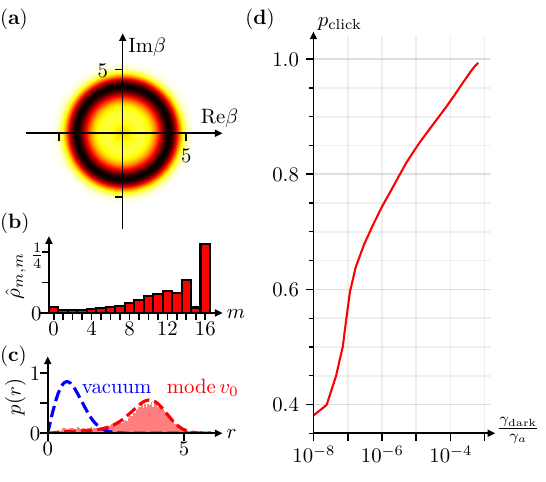}
    \caption{(a) The quantum state $\hat{\rho}$  of the dominant eigenmode for the two-stage \AMP\ device of Fig.~\ref{fig:fig_2stage_modes}(d) yields a ring-shaped Husimi-Q distribution with large $\ex{|\beta|}$ corresponding to (b) large occupations of high Fock states. (c) The resulting histogram of $|\beta|=r$ can  clearly be distinguished from the vacuum distribution (theory curves with dashed lines) and (d) a similar detection scheme as before yields improved detection probabilities. 
    [Parameters: $n_1=n_2=4$, $\sigma_\omega = \gamma_b = \gamma_a / 100$, $\kappa_0=0.2$, $\beta_0 = 1.67$, $\epsilon_{J,1}=0.89 \epsilon_{J,1}^\mathrm{opt}$, $\epsilon_{J,2}=0.1 \epsilon_{J,2}^\mathrm{opt}$, $\tau_c = 60$].}
    \label{fig:fig_2stage_results}
\end{figure}

In the last subsection, despite the more complex two-stage dynamics, we found a regime with dominant single-mode occupation. Since that occupation for the $3\times 3$ amplifier is not quite as high as for the single-stage $n=9$ case, we turn to a $4\times 4$ scenario, where we similarly identify a good regime of $\epsilon_{J,1/2}$ for a finite $\beta_0$.
The Husimi-Q function of the dominant output eigenmode is shown in Fig.~\ref{fig:fig_2stage_results}(a). The corresponding occupation distribution, Fig.~\ref{fig:fig_2stage_results}(b), confirms large occupation of higher Fock-states with a curious (and as yet not understood) suppression of states $\ket{15}$ and $\ket{13}$. Crucially, separability from the vacuum, Fig.~\ref{fig:fig_2stage_results}(c), is good. Proceeding in the same analysis scheme as for the single stage, cf. Sec. \ref{sec::Detection1stage_Scheme1},
we arrive at the performance curve shown in Fig.~\ref{fig:fig_2stage_results}(d) and can, for instance for a photon rate of $\gamma_\mathrm{ph} = 1/(5\tau_\mathrm{c}) = \gamma_a/3000$, detect $84.5$ out of $100$ photons correctly with $3$ dark counts. 
This corresponds to a dark count rate of $\gamma_\mathrm{dark} = 10^{-5}\gamma_a$, or one dark count in a time equivalent to 1000 times the length of the impinging Gaussian pulse.
While the investigated two-stage \AMP\ only slightly outperforms the single-stage device [c.\,f.\, Fig.~\ref{fig:fig_1stage_scheme1_1}(d) and \ref{fig:fig_1stage_scheme1_2}(d)], it uses more easily achievable parameters for Josephson driving strengths and photon multiplication factors ($n_1\times n_2 = 4^2$ versus $n=9$). Moreover, one expects very favorable improvements even upon increasing only to $n_{1/2}=5$, which, however, may already reach the limits of a comprehensive numerical analysis. The parameters assumed in our schemes for zero-point fluctuations, quality factors and Josephson energies are experimentally accessible. While Josephson energies can be tuned in-situ using a SQUID geometry, large zero-point fluctuations such as $\beta_0\approx 1.67$ can be engineered when increasing the inductance-to-capacitance ratio of the resonator (e.g.~zero-point fluctuations of $\mathcal{O}(1)$ have been achieved in \cite{Rolland2019, Menard2022}).

\section{Conclusion}
\label{sec::Conclusion}
We have demonstrated that an inelastic Cooper pair tunneling photon multiplier can be used to detect a single impinging microwave photon.  
In such a system, two microwave modes are connected in-series to a dc-driven Josephson junction which drives a photon multiplication process to convert one photon from mode $\ann$ to $n$ photons in cavity $\bnn$. When the Josephson energy fulfills an impedance-matching condition, a resonant Gaussian photon pulse impinging from the left transmission line is absorbed by cavity $\ann$ and (quasi-)deterministically converted to $n$ photons leaking from cavity $\bnn$ into the right transmission line, where they can be detected by a heterodyne detection scheme which models the experimental scenario of a weak quadrature measurement after quantum-limited amplification of the output signal.
The detection scheme relies not only on a high conversion probability, but on a highly-occupied output mode, whose Husimi-Q function can be well discriminated from vacuum. 
We have therefore characterized the multi-mode output of the \AMP\ in different parameter regimes and identified optimal parameter ranges that ensure both near-perfect photon conversion and high photon occupation in a single output pulse.

Our detection scheme can be modified or extended for other scenarios, like detection and counting of the simultaneous arrival of several input photons or optimized for different expected pulse shapes. 

We have presented a single-stage setup with a multiplication factor of $n\approx 9$, which is experimentally challenging due to unwanted cavity excitations from higher resonances due to the required large dc voltage, as well as a two-stage extension. While in principle three multiplication stages are possible, the two-stage device with multiplication of $n=4\cdot 4$ yields sufficient detection probabilities for low dark counts: 
Realistic simulations of the experimental scenario with an impinging Gaussian pulse of length $T$ yield a photon detection probability of $84.5\%$ with $10^{-3}/T$ dark count rate, which is competitive with existing microwave detectors. 
Combining the \AMP\ with other setups for microwave-photon amplification or augmenting the two stage multiplication to $n>4$ will significantly improve the detection efficiency beyond the  numbers reported here. 
The \AMP\ based on photon number amplification realizes a detector for single itinerant microwave photons without dead times, with low dark counts and detection efficiencies well beyond $80\%$.

\section{Acknowledgements}
\label{sec::Acknowledgements}
JA acknowledges financial support from the German Science Foundation (DFG) through grants AN336/13-1, AN336/18-1; JA, CP, BK through the Center for Integrated Quantum Science and Technology (IQST), and MH and NB from the Natural Sciences and Engineering Research Council of Canada (NSERC) through grants RGPIN-2025-06130, ALLRP 565748-22 and from the Québec government through Prompt Quebec grant 05\_AQ22.001-V3.

\appendix

\begin{widetext}
\section{Hamiltonians in rotating  Rotating-Wave Approximation}

 In this Appendix, we provide the derivation for the rotating-wave Hamiltonians of the \AMP\ devices studied in the main text. 

 \subsection{Single-Stage Amplifier}
 \label{sec::app_1stage_RWA}

The circuit from Fig.~\ref{fig:fig_model}(a), consisting of two microwave cavities connected in series with a dc-biased Josephson junction, can be described by the Hamiltonian 
\begin{align}
    \hat{H}_1 = \hat{H}_\mathrm{res} + \hat{H}_J
\end{align}
 consisting of a free resonator Hamiltonian which is driven by the Josephson junction, 
 \begin{align}
   \hat{H}_\mathrm{res} 
   &=  \hbar \omega_a \adag \ann + \hbar \omega_b \bdag \bnn \\
    \hat{H}_J 
    &= - E_J \cos\left[\omega_\mathrm{dc} t + \alpha_0(\adag + \ann) + \beta_0(\bdag + \bnn) \right]  \, . 
 \end{align}
 We assume that the dc-voltage associated with the frequency $\omega_\mathrm{dc} = 2eV_\mathrm{dc} / \hbar$ is applied at the resonance $\omega_\mathrm{dc} \approx n \omega_b - \omega_a$ and move to a rotating reference frame defined by the unitary transformation $\hat{U} = \hat{U}_a \otimes \hat{U}_b$ with 
\begin{equation}
   \label{eq:app_U_rot_r}
    \hat{U}_r = \exp[i \phi_\mathrm{rot,r} \hat{r}^\dagger \hat{r}] = \exp[i (\omega_\mathrm{rot,r} t + \phi_r) \hat{r}^\dagger \hat{r}]
\end{equation}
for $r=a,b$. Fixing the sum of rotating-frame frequencies to the driving frequency, $\omega_\mathrm{dc} = n \omega_{\mathrm{rot}, b} - \omega_{\mathrm{rot},a}$ transforms the free resonator Hamiltonian to 
\begin{equation}
\label{eq:singlemode_rwa_rot} 
\hat{H}_\mathrm{res, rot} = \hbar (\omega_a - \omega_\mathrm{rot,a}) \adag \ann + \hbar (\omega_b - \omega_\mathrm{rot,b}) \bdag \bnn \, . 
\end{equation}
 Throughout the paper, we assume perfectly resonant driving, such that $\hat{H}_\mathrm{res, rot} = 0$. In the driving Hamiltonian, we now additionally perform a rotating wave approximation (keeping only slowly oscillating terms), which yields  
 \begin{equation}
    \label{eq:app_1stage_H_RWA}
     \hat{H}_{J,\mathrm{RWA}} = \frac{E_J^* \alpha_0 \beta_0^n}{2 n!} :(\ann \bdag[n] + \adag \bnn^n) \frac{J_1\left(2\alpha_0\sqrt{\adag \ann}\right)}{\alpha_0 \sqrt{\adag \ann}} \frac{J_n\left(2\beta_0\sqrt{\bdag \bnn}\right)\cdot n!}{\left(\beta_0 \sqrt{\bdag \bnn}\right)^n}: 
\end{equation}
Here, $E_J^* = E_J \cdot e^{-(\alpha_0^2 + \beta_0^2)/2}$ is the renormalized Josephson energy, $J_n$ are Bessel functions of the first kind, and the colons signal normal ordering of operators. The Hamiltonian contains only nonzero matrix elements in the entries $H_{ m_a+1, m_b |m_a, m_b+n}^*  = H_{m_a, m_b+n | m_a+1, m_b} = \bra{m_a}\bra{m_b+n} \hat{H}_{J,\mathrm{RWA}} \ket{m_a +1} \ket{m_b}$ (with $m_a, m_b \in \mathbb{N}_0$), where 
\begin{align}
 H_{m_a, m_b+n | m_a+1, m_b} = \frac{E_J^* \alpha_0 \beta_0^n}{2 n!} \left[\sqrt{m_a+1} \frac{L_{m_a}^{(1)}(\alpha_0^2)}{L_{m_a}^{(1)}(0)} \right]\cdot \left[\sqrt{\frac{(m_b+n)!}{(m_b)!}} \frac{L_{m_b}^{(n)}(\beta_0^2)}{L_{m_b}^{(n)}(0)} \right]  
\end{align}
 and $L_m^{(n)}$ are generalized Laguerre polynomials. For a single incoming photon pulse, the Hamiltonian dimension restricts $m_a = m_b = 0$, yielding only two nonzero matrix elements
 \begin{align}
     H_{0, n | 1, 0 } =H_{1, 0 | 0, n }^* =  \frac{E_J^* \alpha_0 \beta_0^n}{2 n!} \bra{0_a} \ann \ket{1_a} \bra{n_b} \bdag[n] \ket{0_b} =  \frac{E_J^* \alpha_0 \beta_0^n}{2 \sqrt{n!}} \, . 
 \end{align}
 Since only the lowest levels couple, we see that the fractions of Bessel functions in \eqref{eq:app_1stage_H_RWA} are not involved, which yields Hamiltonian \eqref{eq::1stage_Hamiltonian_RWA} from the main text. 

 \subsection{Two-Stage Amplifier}
 \label{sec::app_2stage_RWA}

 Starting now at the Hamiltonian $\hat{H}_2 = \hat{H}_\mathrm{res} + \hat{H}_{J1} + \hat{H}_{J2}$, where now 
 \begin{align}
    \hat{H}_\mathrm{res} 
   &=  \hbar \omega_a \adag \ann + \hbar \omega_b \bdag \bnn + \hbar \omega_c \cdag \cnn \\
    \hat{H}_{J1} 
    &= - E_{J1} \cos\left[\omega_\mathrm{dc,1} t + \alpha_0(\adag + \ann) + \kappa_0(\cdag + \cnn) \right] \\
    \hat{H}_{J2} 
    &= - E_{J2} \cos\left[\omega_\mathrm{dc,2} t + \kappa_0(\cdag + \cnn) + \beta_0(\bdag + \bnn) \right]  \, .      
 \end{align}
 and $\omega_{\mathrm{dc}, i} = 2e V_{\mathrm{dc},i} / \hbar$ such that $\omega_{\mathrm{dc},1} = n_1 \omega_c - \omega_a$ and $\omega_{\mathrm{dc},2} = n_2 \omega_b - \omega_c$, an analogous unitary transformation $\hat{U} = \hat{U}_a \otimes \hat{U}_b \otimes \hat{U}_c$ (with $\hat{U}_r$ defined as in Eq. \ref{eq:app_U_rot_r}) yields $\hat{H}_\mathrm{res} = 0$ on resonance. The rotating-wave approximation then yields 
 
 \begin{equation}
    \begin{split}
     \hat{H}_{\mathrm{RWA}, J1} 
     &= \frac{E_{J,1}^* \alpha_0 \kappa_0^{n_1}}{2 n_1!} :\ann \cdag[n_1] \frac{J_1\left((2\alpha_0\sqrt{\adag \ann}\right)}{\alpha_0 \sqrt{\adag \ann}} \frac{J_{n_1}\left(2\kappa_0\sqrt{\cdag \cnn}\right)\cdot n_1!}{\left(\kappa_0 \sqrt{\cdag \cnn}\right)^{n_1}}: + \mathrm{h.c.}\\
    \hat{H}_{\mathrm{RWA}, J2} 
    &=  \frac{E_{J,2}\kappa_0 \beta_0^{n_2}}{2 n_2!} :\cnn \bdag[n_2] \frac{J_1\left(2\kappa_0\sqrt{\cdag \cnn}\right)}{\kappa_0 \sqrt{\cdag \cnn}} \frac{J_{n_2}\left(2\beta_0\sqrt{\bdag \bnn}\right)\cdot n_2!}{\left(\beta_0 \sqrt{\bdag \bnn}\right)^{n_2}}:  + \mathrm{h.c.}
    \end{split} 
 \end{equation}
  with $E_{J1}^* = E_{J1} e^{-(\alpha_0^2 + \kappa_0^2)/2}$ and $E_{J2}^* = E_{J2} e^{-(\kappa_0^2 + \beta_0^2)/2}$. Here, nonlinear corrections by the Bessel functions must always included, modifying the Hamiltonian transition matrix entries. They are
  \begin{align}
   \label{eq::2stage_laguerre_1}
     H_{m_a, m_b, m_c+n_1 | m_a+1, m_b, m_c} = \frac{E_{J1}^* \alpha_0 \kappa_0^{n_1}}{2 n_1!} \left[\sqrt{m_a+1} \frac{L_{m_a}^{(1)}(\alpha_0^2)}{L_{m_a}^{(1)}(0)} \right]\cdot \left[\sqrt{\frac{(m_c+n_1)!}{(m_c)!}} \frac{L_{m_c}^{(n_1)}(\kappa_0^2)}{L_{m_c}^{(n_1)}(0)} \right]  \\
    \label{eq::2stage_laguerre_2}
     H_{m_a, m_b+n_2, m_c |m_a, m_b, m_c+1} = \frac{E_{J2}^* \kappa_0 \beta_0^{n2}}{2 n_2!} \left[\sqrt{m_c+1} \frac{L_{m_c}^{(1)}(\kappa_0^2)}{L_{m_c}^{(1)}(0)} \right]\cdot \left[\sqrt{\frac{(m_b+n_2)!}{(m_b)!}} \frac{L_{m_b}^{(n_2)}(\beta_0^2)}{L_{m_b}^{(n_2)}(0)} \right]  
\end{align}

\begin{figure}
    \centering
    \includegraphics[width=\columnwidth]{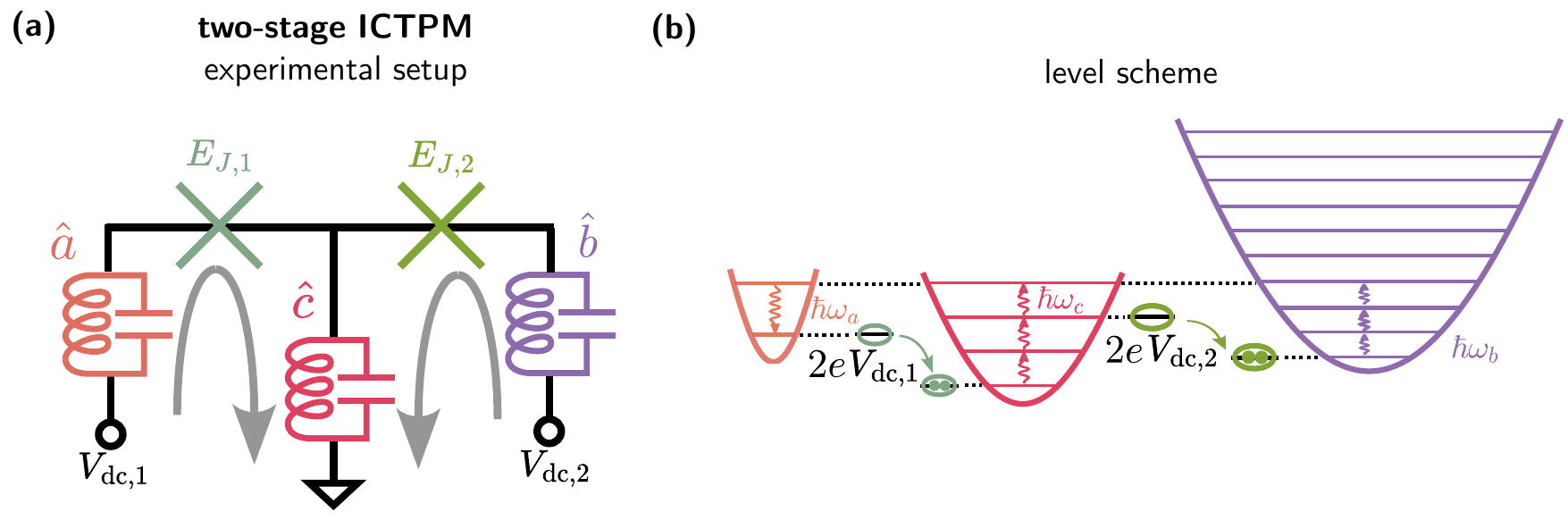}
    \caption{Two-stage \AMP\ as cascaded extension of the single-stage \AMP. (a) In the first circuit loop, the microwave resonators $\ann$ and $\cnn$ are connected in series to a Josephson junction with Josephson energy $E_{J,1}$. The second circuit loop couples the resonators $\cnn$ and $\bnn$ with the Josephson junction with energy $E_{J,2}$. Both circuits are dc-voltage biased and cavity $\ann$ ($\bnn$) are coupled to an input (output) transmission line. (b) Setting the dc voltage in the first circuit to the resonance $2eV_{\mathrm{dc},1} \approx n_1 \omega_c - \omega_a$ will result in photon multiplication of one cavity-$\ann$ photon and $n_1$ cavity-$\cnn$ photons. A second multiplication stage between cavity $\cnn$ and $\bnn$ (with the resonance condition $eV_{\mathrm{dc},2} \approx n_2 \omega_b - \omega_c$) will the ideally yield in total $n_1 \cdot n_2$ (here shown for $3\cdot3=9$) output photons, when appropriately choosing the Josephson energies to achieve perfect photon conversion. }
    \label{fig:app_2stage}
\end{figure}

 \end{widetext}

\section{Cascaded Master Equation Including Quantum Pulses}
 \label{sec::app_cascadedME}

  The Lindblad master equation, that includes quantum pulses modeled as auxiliary cavities, is 
  \begin{align}
      \frac{d \hat{\rho}}{dt} = - \frac{i}{\hbar} [\hat{H}_\mathrm{tot}, \hat{\rho}] + \mathcal{L}_a[\hat{\rho}] + \mathcal{L}_b[\hat{\rho}] \, . 
  \end{align}
   The dissipation (for $i=a,b$) exhibits the standard Lindblad form 
   \begin{align}
       \mathcal{L}_i[\hat{\rho}] = \hat{L}_i \hat{\rho} \hat{L}_i^\dag - \frac{\hat{L}_i^\dag\hat{L}_i \hat{\rho} + \hat{\rho}\hat{L}_i^\dag\hat{L}_i }{2} \, . 
   \end{align}
   The total Hamiltonian $\hat{H}_\mathrm{tot} = \hat{H}_\mathrm{sys} + \hat{H}_\mathrm{int}$ now consists of the system Hamiltonian and coherent interactions
   \begin{align}
       \hat{H}_\mathrm{int}   
       = &\frac{i\hbar}{2} \left[\sqrt{\gamma_a} g_{u_a} \ann_u^\dag \ann + \sqrt{\gamma_a} g_{v_a}^* \adag \ann_v + g_{u_a} g_{v_a}^* \adag_u \ann_v - \mathrm{h.c.}\right] \nonumber \\
       + &\frac{i\hbar}{2} \left[\sqrt{\gamma_b} g_{u_b} \bnn_u^\dag \bnn + \sqrt{\gamma_b} g_{v_b}^* \bdag \bnn_v + g_{u_b} g_{v_b}^* \bdag_u \bnn_v -\mathrm{h.c.} \right] \nonumber \\ \label{eq:app_cascaded_H}
   \end{align}
   between the auxiliary cavities and the system. The Hamiltonian together with now modified loss-operators 
   \begin{align}
       \hat{L}_a &= \sqrt{\gamma_a} \ann + g_{u_a}^* \ann_u + g_{v_a}^* \ann_v \\
       \hat{L}_b &= \sqrt{\gamma_b} \bnn + g_{u_b}^* \bnn_u + g_{v_b}^* \bnn_v
   \end{align}
   ensure in a cascaded manner \cite{Kiilerich2019,Kiilerich2020, GardinerQuantumNoise} a directional influence between the subsystems: The input cavities $\ann_u$ and $\bnn_u$ influence the system's time evolution, but their time-evolution are not influenced by the system or the output cavities. Likewise, the time-evolution of the system is not influenced by the output cavities $\ann_v$ and $\bnn_v$, but their time evolution is influenced by the input cavities and the system. The auxiliary input (output) cavities have mathematically constructed time-dependent loss rates, defined as 
   \begin{align}
       g_u(t) &= \frac{u^*(t)}{\sqrt{1 - \int_0^t |u(\tau)|^2 d\tau}} \\
       g_v(t) &= -\frac{v^*(t)}{\sqrt{\int_0^t |v(\tau)|^2 d\tau}} \, . 
   \end{align}
    In this way, they only emit (absorb) a quantum pulse given as initial (final) state of the specific mode envelope $u(t)$ ($v(t)$). A generalization of Molmer's approach involving multiple auxiliary cavities can be found in Ref. \cite{Kiilerich2020}. In this paper, we typically consider a Gaussian single-photon pulse into the system through the input (left) transmission line and do not specifically model the reflected output from cavity $\ann$ (into the left transmission line), which here is formally achieved by setting $g_{v_a} = 0$ and by choosing $u_a(t)$ as a Gaussian mode envelope. Secondly, we always consider vacuum input into the system from the output (right) transmission line ($g_{u_b}=0$). If we want to find the quantum state of a specific output mode $v_b(t)$, we include it with a accordingly defined $g_{v_b}(t)$, else we also set $g_{v_b}(t)=0$.

\section*{References}
    
\bibliographystyle{ieeetr}
\bibliography{references}

\end{document}